\documentclass[a4paper,fleqn,usenatbib]{mnras}

\usepackage[T1]{fontenc}
\usepackage{ae,aecompl}

\usepackage{graphicx}	
\usepackage{amsmath}	
\usepackage{amssymb}	
\usepackage{mathrsfs}                
\usepackage{booktabs}
\usepackage{threeparttable}

\title[Constraints on LIV with Polarized GRBs]{New Constraints on Lorentz Invariance Violation with Polarized Gamma-Ray Bursts}

\author[Wei]{
Jun-Jie Wei$^{1,2}$\thanks{E-mail: jjwei@pmo.ac.cn (JJW)}
\\
$^{1}$Purple Mountain Observatory, Chinese Academy of Sciences, Nanjing 210034, China\\
$^{2}$Guangxi Key Laboratory for Relativistic Astrophysics, Guangxi University, Nanning 530004, China
}

\date{Accepted XXX. Received YYY; in original form ZZZ}

\pubyear{2018}

\begin{document}
\label{firstpage}
\pagerange{\pageref{firstpage}--\pageref{lastpage}}
\maketitle

\begin{abstract}
Lorentz invariance violations (LIV) can yield vacuum birefringence, which results in an energy-dependent rotation
of the polarization vector of linearly polarized emission from astrophysical sources. It is believed that if the relative
rotation angle ($\Delta\Theta$) of the polarization vector of high energy photons with respect to that of low energy
photons is larger than $\pi/2$, then the net polarization of the signal would be significantly depleted and could not
be as high as the observed level. Hence, the measurement of high polarization implies that $\Delta\Theta$ should not
be too large. In this work, we assemble recent detections of prompt emission polarization in gamma-ray bursts (GRBs),
all of whom have high detection significance. Following the method shown in \cite{2016MNRAS.463..375L},
we give a detailed calculation on the polarization evolution arising from the LIV effect for each GRB,
and confirm that, even if $\Delta\Theta$ is approaching to $\pi/2$, the net polarization is not severely suppressed,
and more than 60\% of the initial polarization can be conserved. Applying this method to our GRB polarimetric
data, we improve existing sensitivities to LIV involving photons by factors ranging from two to ten.
In addition, we prove that our constraints are not greatly affected by uncertainties in the spectral parameters of GRBs.
\end{abstract}

\begin{keywords}
astroparticle physics -- gravitation -- polarization -- gamma-ray burst: general
\end{keywords}



\section{Introduction}
\label{sec:Intro}
Lorentz invariance is the fundamental symmetry of Einstein's relativity.
However, various quantum gravity theories predict that Lorentz symmetry may be broken
at the Planck energy scale $M_{\rm Pl}\simeq1.22\times10^{19}$ GeV
\citep{1989PhRvD..39..683K,1991NuPhB.359..545K,1995PhRvD..51.3923K,1998Natur.393..763A,2005LRR.....8....5M,2005hep.ph....6054B,2013LRR....16....5A,2014RPPh...77f2901T}.
Experimental searches for deviations from Lorentz invariance have been performed in
a wide range of systems (see \citealt{2011RvMP...83...11K} for a compilation).

In the photon sector, signatures of Lorentz invariance violations (LIV) include
vacuum dispersion and vacuum birefringence \citep{2008ApJ...689L...1K}.
Even though these effects are expected to be very tiny at observable energies
$\ll M_{\rm Pl}$, they can accumulate over large distances and become detectable.
Astronomical measurements involving the long baselines can therefore provide sensitive tests of Lorentz symmetry.
Vacuum dispersion causes an energy-dependent speed of light. Hence, the arrival-time differences of photons
with different energies emitted simultaneously from an astrophysical source can be used to test Lorentz invariance
(e.g.,
\citealt{1998Natur.393..763A,
2005PhLB..625...13P,
2006APh....25..402E,
2008JCAP...01..031J,
2008ApJ...689L...1K,
2009PhRvD..80a5020K,
2009Sci...323.1688A,
2012APh....36...47C,
2012PhRvL.108w1103N,
2013PhRvD..87l2001V,
2013APh....43...50E,
2015PhRvD..92d5016K,
2015APh....61..108Z,
2017ApJ...834L..13W,
2017ApJ...842..115W,
2017ApJ...851..127W,
Liu2018}).
Similarly, vacuum birefringence leads to a wavelength-dependent rotation of the polarization
vector of a linearly polarized light. Lorentz invariance can therefore be tested through
polarimetric observations of astrophysical sources
(e.g.,
\citealt{1999PhRvD..59l4021G,
2001PhRvD..64h3007G,
2001PhRvL..87y1304K,
2006PhRvL..97n0401K,
2007PhRvL..99a1601K,
2013PhRvL.110t1601K,
2003Natur.426Q.139M,
2004PhRvL..93b1101J,
2007MNRAS.376.1857F,
2009JCAP...08..021G,
2011PhRvD..83l1301L,
2011APh....35...95S,
2012PhRvL.109x1104T,
2013MNRAS.431.3550G,
2014MNRAS.444.2776G,
2016MNRAS.463..375L,
2017PhRvD..95h3013K,
2018arXiv180908356F}).
Typically, more stringent limits on LIV result from polarization measurements
rather than vacuum dispersion measurements.
This can be understood by the fact that the former are more
sensitive than the latter by a factor $\propto 1/\omega$,
where $\omega$ is the frequency of the observed light \citep{2009PhRvD..80a5020K}.

Owing to the high energy polarimetry of prompt emission and their large cosmological distances,
gamma-ray bursts (GRBs) have been viewed as promising sources for searching for
LIV-induced vacuum birefringence
\citep{
2003Natur.426Q.139M,
2004PhRvL..93b1101J,
2006PhRvL..97n0401K,
2007PhRvL..99a1601K,
2013PhRvL.110t1601K,
2007MNRAS.376.1857F,
2011PhRvD..83l1301L,
2011APh....35...95S,
2012PhRvL.109x1104T,
2013MNRAS.431.3550G,
2014MNRAS.444.2776G,
2016MNRAS.463..375L}.
If Lorentz invariance is broken then group velocities of photons with right- and left-handed circular polarizations
should differ slightly, leading to a phase rotation of linear polarization.
Observations of linear polarization can therefore place strict limits on the birefringence parameter
($\xi$). It is believed that if high polarization is observed, then the difference of rotation angles (denoted by
$\Delta\Theta$) of polarization vectors between the highest and lowest energy photons should not be too large.
Otherwise, the net polarization of the signal will be significantly depleted and could not be as high as the observed level.
\cite{2012PhRvL.109x1104T} set the upper limit of the relative rotation angle $\Delta\Theta$ to be $\pi/2$, and
obtained a severe upper bound on the birefringence parameter $\xi$ in the order of $\mathcal{O}(10^{-15})$ from
the polarimetric data of GRB 110721A. However, GRB 110721A had no direct redshift measurement at that time,
they used a distance estimate based on the empirical luminosity relation. By using the redshift determination
($z=1.33$) together with the polarization measurement of GRB 061122, \cite{2013MNRAS.431.3550G} obtained a
much stricter limit, $\xi<3.4\times10^{-16}$. The most distant polarized burst, GRB 140206A, which has a confirmed
redshift measurement of $z=2.739$, yielded the deepest limit to date
\citep{2014MNRAS.444.2776G}. All these limits were based on the assumption that the differential rotation angle
$\Delta\Theta$ is smaller than $\pi/2$. However, \cite{2016MNRAS.463..375L} calculated the evolution of GRB
polarization arising from the LIV effect, and found that more than 60\% of
the initial polarization (depending both on the photon spectrum and the photon energy range) can still be conserved
at $\Delta\Theta\leq\pi/2$.
It is interesting to constrain the LIV effect under the framework of calculating GRB polarization evolution.

In this paper, following the calculation proposed in \cite{2016MNRAS.463..375L} and using the latest measurements
of linear polarization in GRBs, we update constraints on a possible deviation from Lorentz invariance through
the vacuum birefringence effect, and thereby improve existing sensitivities to LIV involving photons
by factors ranging from two to ten. Moreover, since the evolution of polarization induced by the LIV effect is related
to the GRB spectrum \citep{2016MNRAS.463..375L}, we also investigate the impact of the observational uncertainties
of spectral parameters on the resulting constraints.
The paper is organized as follows. In Section~\ref{sec:formulae}, we give an overview of the general formulae
for the polarization evolution of photons. The polarization sample at our disposal and the corresponding
constraint results are presented in Sections~\ref{sec:data} and \ref{sec:results}, respectively.
Finally, we draw a brief summary and discussion in Section~\ref{sec:summary}. Throughout we adopt the unit $\hbar=c=1$.

\section{General Formulae}
\label{sec:formulae}
The Lorentz-violating dispersion relation for photon propagation takes the generalized form \citep{2003PhRvL..90u1601M}
\begin{equation}\label{eq:dispersion}
  E_{\pm}^2=p^2\pm \frac{2\xi}{M_{\rm pl}} p^3\;,
\end{equation}
where $M_{\rm pl}\approx 1.22\times 10^{19}$ GeV is the Planck energy, $\pm$ represents different circular polarization states,
and $\xi$ is a dimensionless parameter characterizing the order of magnitude of the LIV effect. If $\xi\neq0$, then
the dispersion relation will lead to slightly different propagation velocities for different polarization states.
Thus, the polarization vector of a linearly polarized light will rotate during the propagation. This effect
is known as vacuum birefringence.
The rotation angle during the propagation from the source at redshift $z$ to the observer
is expressed as \citep{2011PhRvD..83l1301L,2012PhRvL.109x1104T}
\begin{equation}\label{eq:delta-theta}
  \Delta\theta(k)=\xi\frac{k^2}{M_{\rm pl}H_0}\int_0^z\frac{(1+z')dz'}{\sqrt{\Omega_{\rm m}(1+z')^3+\Omega_{\Lambda}}}\;,
\end{equation}
where $k$ is the energy of the observed light.
Here we use the standard cosmological parameters:
$H_{0}=67.3$ km $\rm s^{-1}$ $\rm Mpc^{-1}$, $\Omega_{\rm m}=0.315$, and $\Omega_{\Lambda}=0.685$
\citep{2014A&A...571A..16P}.

Following the process of calculating the polarization evolution in \citet{2016MNRAS.463..375L},
we assume that there is a beam of non-coherent light emitted from a source, in which
the propagation direction is selected as the $z$-axis and the polarization direction is in the $xy$-plane.
The intensity of photons with electric vector in the infinitesimal azimuth angle interval $d\theta$ and
with energy in the infinitesimal interval $dk$ is given by
\begin{equation}\label{eq:intensity1}
  dj(\theta,k)=j_0f(\theta)kN(k)d\theta dk\;,
\end{equation}
where $j_0$ denotes a normalized constant, $N(k)$ is the photon spectrum, and $f(\theta)$ represents
a periodic function of $\theta$ with period $\pi$.
Because of the positive correlation between the photon intensity and the square of the electric vector,
the intensity projected onto the direction of azimuth angle $\varphi$ can be written as
\begin{equation}\label{eq:intensity2}
  dj_{\varphi}(\theta,k)=j_0f(\theta)kN(k)\cos^2(\varphi-\theta)d\theta dk\;.
\end{equation}
The initial total intensity of photons polarized along the direction $\varphi$ is then expressed as
\begin{equation}\label{eq:intensity3}
  I(\varphi)=\int dj_{\varphi}(\theta,k)=\int_0^{\pi}d\theta\int_{k_1}^{k_2}dk~j_0f(\theta)kN(k)\cos^2(\varphi-\theta)\;,
\end{equation}
where $k_1<k<k_2$ is the energy range of photon spectrum.
The polarization degree is defined as \citep{1979rpa..book.....R}
\begin{equation}\label{eq:polarization1}
  \Pi=\frac{I_{\rm max}-I_{\rm min}}{I_{\rm max}+I_{\rm min}}\;,
\end{equation}
where $I_{\rm max}$ and $I_{\rm min}$ correspond to the maximum and minimum values of $I(\varphi)$, respectively.

As most GRBs have a relatively large level of polarization \citep{2017NewAR..76....1M},
we consider that the photon beam is initially completely polarized, e.g., along the $x$-axis.
To ensure $f(\theta)$ is periodic, i.e., $f(\theta+\pi)=f(\theta)$,
the sum of $\delta$-functions is adopted \citep{2016MNRAS.463..375L}:
\begin{equation}\label{eq:f-theta}
  f(\theta)=\sum_{n=-\infty}^{+\infty}\delta(\theta-n\pi)\;.
\end{equation}
Substituting Equation (\ref{eq:f-theta}) into Equation (\ref{eq:intensity3}),
the initial photon intensity simplifies to
\begin{equation}
  I(\varphi)=j_0\cos^2\varphi \int_{k_1}^{k_2}dk~kN(k)\;.
\end{equation}
It is obvious that $I_{\rm min}=I(\pi/2)=0$ and $I_{\rm max}=I(0)$, which implies that the initial polarization degree $\Pi$
is $100\%$, i.e., the initial polarization state is initially completely polarized.

Vacuum birefringence results in a rotation of the plane of linear polarization for photons at different energies
emitted with the same initial polarization degree. Let $\Delta\Theta\equiv\Delta\theta(k_2)-\Delta\theta(k_1)$
to be the relative rotation angle of the polarization vector of high energy photons with respect to that
of low energy photons, then Equation (\ref{eq:delta-theta}) can be rephrased as
\begin{equation}\label{eq:delta-theta2}
  \Delta\theta(k)=\Delta\Theta\frac{k^{2}}{k_2^{2}-k_1^{2}}\;.
\end{equation}
The received photon intensity can then be derived by replacing $f(\theta)$ with $f(\theta+\Delta\theta(k))$
in Equation (\ref{eq:intensity3}), i.e.,
\begin{equation}\label{eq:intensity4}
  I'(\varphi)=\int_0^{\pi}d\theta\int_{k_1}^{k_2}dk~j_0f\left(\theta+\Delta\theta(k)\right)kN(k)\cos^2(\varphi-\theta)\;.
\end{equation}
Substituting Equations (\ref{eq:f-theta}) and (\ref{eq:delta-theta2})
into Equation (\ref{eq:intensity4}), we have the observed photon intensity
\begin{equation}
  I'(\varphi)
  =j_0 \int_{k_1}^{k_2}dk~kN(k)\cos^2\left(\varphi+\frac{\Delta\Theta k^{2}}{k_2^{2}-k_1^{2}}\right)\;.
  \label{eq:observed-intensity}
\end{equation}
Accordingly, the observed polarization degree is
\begin{equation}\label{eq:polarization2}
  \Pi'=\frac{I'_{\rm max}-I'_{\rm min}}{I'_{\rm max}+I'_{\rm min}}\;,
\end{equation}
where $I'_{\rm max}$ and $I'_{\rm min}$ are the maximum and minimum values of $I'(\varphi)$, respectively.

\section{Polarization measurements of prompt GRB emission}
\label{sec:data}

\begin{table*}
\caption{Limits on LIV from GRB polarization measurements}
\begin{center}
\begin{tabular}{lccccccccc}
\hline
\hline
GRB & Instrument & $\rm Energy$ &    $\Pi$    & \multicolumn{3}{c}{Spectral parameters (90\% C.L.)} & \emph{z} & $\rm Refs.^{b}$ & $\xi$\\
\cmidrule(lr){5-7}
    &            & (keV)  & (68\% C.L.) &    $\alpha$  & $\beta$  & $E_{p}/E_{c}$               &          &       &  (68\% C.L.)    \\
    &            &        &             &             &         & (keV)                       &          &       &      \\
\hline
061122	& INTEGRAL/IBIS	&	250---800	&$	>60\%	$&	$	-1.15	^{+	0.04}_{-0.04}$	&	---	&	$	221	^{+	20}_{-20}$	&	1.33	& 1, 1, 1	&	$<5.2\times10^{-17}$\\
100826A	& IKAROS/GAP	&	70---300	&$	27\pm11\%	$&	$	-1.31^{+0.06}_{-0.05}$	&	$	-2.10^{+0.10}_{-0.20}$	&	$	606	^{+	134}_{-109}$	&	$\rm 0.083^{a}$	& 2, 3, 4	&	$1.2^{+1.4}_{-0.7}\times10^{-14}$\\
110301A	& IKAROS/GAP	&	70---300	&$	70\pm22\%	$&	$	-0.81^{+0.02}_{-0.02}$	&	$	-2.70^{+0.04}_{-0.05}$	&	$	106.8^{+1.85}_{-1.75}$	&	$\rm 0.082^{a}$	& 5, 6, 4	&	$4.3^{+5.4}_{-2.3}\times10^{-15}$\\
110721A	& IKAROS/GAP	&	70---300	&$	84^{+16}_{-28}\%$&	$	-0.94^{+0.02}_{-0.02}$	&	$	-1.77^{+0.02}_{-0.02}$	&	$	372.5^{+	26.5}_{-23.6}$	&	0.382	& 5, 7, 8	&	$5.1^{+4.0}_{-5.1}\times10^{-16}$\\
140206A	& INTEGRAL/IBIS	&	200---400	&$	>48\%	$&	$	1.10^{+	0.15}_{-0.15}$	&	---	&	$	114	^{+	47}_{-26}$	&	2.739	& 9, 9, 9	&	$<1.0\times10^{-16}$\\
160106A	& AstroSat/CZTI	&	100---300	&$	68.5	\pm	24\%	$&	$	-0.53^{+0.07}_{-0.06}$	&	$	-2.31^{+0.14}_{-0.21}$	&	$	400	^{+	45}_{-40}$	&	$\rm 0.091^{a}$	& 10, 10, 4	&	$3.4^{+1.4}_{-1.8}\times10^{-15}$\\
160131A	& AstroSat/CZTI	&	100---300	&$	94	\pm	31\%	$&	$	-1.16^{+0.04}_{-0.04}$	&	$	-1.56^{+0.07}_{-0.10}$	&	$	586	^{+	518	}_{-259	}$	&	0.972	& 10, 10, 11	&	$1.2^{+2.0}_{-1.2}\times10^{-16}$\\
160325A	& AstroSat/CZTI	&	100---300	&$	58.75\pm23.5\%	$&	$	-0.71^{+0.07}_{-0.06}$	&	$	-2.26^{+0.20}_{-0.30}$	&	$	238	^{+	25	}_{-22	}$	&	$\rm 0.161^{a}$	& 10, 10, 4	&	$2.3^{+1.0}_{-0.9}\times10^{-15}$\\
160509A	& AstroSat/CZTI	&	100---300	&$	96	\pm	40\%	$&	$	-0.75^{+0.02}_{-0.02}$	&	$	-2.13^{+0.03}_{-0.03}$	&	$	334	^{+	12	}_{-10	}$	&	1.17	& 10, 10, 12	&	$0.8^{+2.2}_{-0.8}\times10^{-16}$\\
160802A	& AstroSat/CZTI	&	100---300	&$	85	\pm	29\%	$&	$	-0.61^{+0.04}_{-0.04}$	&	$	-2.40^{+0.10}_{-0.13}$	&	$	280	^{+	17	}_{-14	}$	&	$\rm 0.105^{a}$	& 10, 10, 4	&	$2.0^{+1.7}_{-2.0}\times10^{-15}$\\
160821A	& AstroSat/CZTI	&	100---300	&$	48.7\pm	14.6\%	$&	$	-0.97^{+0.01}_{-0.01}$	&	$	-2.25^{+0.03}_{-0.03}$	&	$	866	^{+	25	}_{-24	}$	&	$\rm 0.047^{a}$	& 10, 10, 4	&	$8.9^{+1.7}_{-1.7}\times10^{-15}$\\
160910A	& AstroSat/CZTI	&	100---300	&$	93.7\pm	30.92\%	$&	$	-0.36^{+0.03}_{-0.03}$	&	$	-2.38^{+0.05}_{-0.06}$	&	$	330	^{+	13	}_{-13	}$	&	$\rm 0.272^{a}$	& 10, 10, 4	&	$4.7^{+7.6}_{-4.7}\times10^{-16}$\\
\hline
\end{tabular}
\end{center}
\begin{tablenotes}
\footnotesize
\item (a) The redshifts of the seven GRBs are estimated by the luminosity relation.
\item (b) The references appear in the following order: polarization, spectral parameters, and redshift: [1] \cite{2013MNRAS.431.3550G}; [2] \cite{2011ApJ...743L..30Y};
[3] \cite{2010GCN.11158....1G}; [4] This work; [5] \cite{2012ApJ...758L...1Y}; [6] \cite{2011GCN.11771....1F}; [7] \cite{2011GCN.12187....1T}; [8] \cite{2011GCN.12193....1B};
[9] \cite{2014MNRAS.444.2776G}; [10] \cite{2017arXiv170706595C}; [11] \cite{2016GCN.18966....1D}; [12] \cite{2016GCN.19419....1T}.
\end{tablenotes}
\label{tab:1}
\end{table*}

The measurement of polarization in the prompt gamma-ray emission of GRBs has always been challenging (see \citealt{2016A&AT...29..205C,2017NewAR..76....1M}
for reviews). The first reported highly polarized burst was GRB 021206, with a linear polarization degree of
$\Pi=80\pm20\%$ \citep{2003Natur.423..415C}. However, subsequent re-analyses of the same data could not confirm
significant polarization signal \citep{2004MNRAS.350.1288R,2004ApJ...613.1088W}. Another early attempt to measure
linear polarization during the prompt phase of GRBs was performed by \cite{2005A&A...439..245W}, who reported evidence
of high polarization in GRB 930131 and GRB 960924, $\Pi>35\%$ and $\Pi>50\%$, respectively. But unfortunately
the authors' estimation method did not allow them to statistically constrain such results, but called for
further independent measurements to confirm whether these two bursts are highly polarized. \cite{2007ApJS..169...75K}
reported a high level of polarization for GRB 041219A ($\Pi=98\pm33\%$), but again this result was criticized
by more detailed analyses \citep{2007A&A...466..895M,2009ApJ...695L.208G}.

Fortunately, evidence of polarized gamma-ray emission from GRBs has been accumulated since 2011.
The gamma-ray burst polarimeter (GAP) onboard the Interplanetary Kite-craft Accelerated by Radiation Of the Sun (IKAROS)
detected a polarization signal with a polarization degree of $\Pi=27\pm11\%$ from GRB 100826A, and a null polarization degree
is ruled out with $2.9\sigma$ confidence level (C.L.) \citep{2011ApJ...743L..30Y}.
IKAROS/GAP also detected gamma-ray polarizations of two other bursts with high significance levels,
with $\Pi=70\pm22\%$ for GRB 110301A, and $\Pi=84^{+16}_{-28}\%$ for GRB 110721A \citep{2012ApJ...758L...1Y}.
The detection significance is $3.7\sigma$ and $3.3\sigma$, respectively.
Using a different instrument, the Imager on Board the International Gamma-Ray Astrophysics Laboratory
(INTEGRAL) Satellite (IBIS), \cite{2013MNRAS.431.3550G} reported a polarization measurement in the prompt
emission of GRB 061122. They put a lower limit on its polarization degree of $\Pi>60\%$ at $68\%$ C.L..
One other highly polarized burst (GRB 140206A) with the polarization degree of $\Pi>48\%$ at $68\%$ C.L.
was also detected by INTEGRAL/IBIS \citep{2014MNRAS.444.2776G}. Recently, \cite{2017arXiv170706595C} presented
the polarization data for the brightest 11 bursts detected by CZTI onboard AstroSat during the first year of
operation. They found that only 7 of these show clear polarization signatures with $\geq3\sigma$ detection significance
for 4 bursts (GRB 160131A, GRB 160802A, GRB 160821A, and GRB 160910A) and $\sim2.5\sigma$ significance
for another 3 bursts (GRB 160106A, GRB 160325A, and GRB 160509A).\footnote{Note that the polarization measurements
of the remaining four bursts (GRB 151006A, GRB 160607A, GRB 160623A, and GRB 160703A) with low significance levels
are not included in our sample.} All of these detections are evidences that the prompt emission of GRBs are highly polarized.
Therefore, the reliable observations of gamma-ray linear polarization mentioned here enable us to obtain
strict limits on LIV.

In assembling the GRB sample, we require the members have an independent polarization detection with high level of significance.
We exclude those bursts whose polarization measurements remain under debate (e.g., GRB 930131, GRB 960924, GRB 021206, GRB 041219A).
Lastly, our sample includes 12 GRBs for which strong evidence of gamma-ray linear polarization exists.
All of these data are obtained from previously published studies, which are shown in Table~\ref{tab:1}.
The first eight columns include the following information for each GRB: (1) the burst name; (2) the instrumentation
for gamma-ray polarimetry; (3) the energy rang in which polarization is observed; (4) the observed polarization degree $\Pi$
(with the corresponding $68\%$ C.L. uncertainty); the spectral parameters (with $90\%$ C.L. uncertainties),
including (5) the low-energy photon index $\alpha$, (6) the high-energy photon index $\beta$,
(7) the spectral peak energy $E_{p}$ or the breaking energy $E_{c}$; and (8) the redshift.

Of these 12 GRBs, five GRBs have redshift measurements, while others have none.
The empirical luminosity relation (the well-known Amati relation; \citealt{2002A&A...390...81A})
is therefore applied to estimate the redshifts of the other seven GRBs. We use the
observed fluence and $E_{p}$ of the seven bursts
(GRB 100826A: 20--10000 keV fluence $3.0\times10^{-4}$ erg $\rm cm^{-2}$ and $E_{p}=606$ keV \citep{2010GCN.11158....1G};
GRB 110301A: 10--1000 keV fluence $3.65\times10^{-5}$ erg $\rm cm^{-2}$ and $E_{p}=106.8$ keV \citep{2011GCN.11771....1F};
GRB 160106A: 100--300 keV fluence $3.5\times10^{-5}$ erg $\rm cm^{-2}$ and $E_{p}=400$ keV;
GRB 160325A: 100--300 keV fluence $7.6\times10^{-6}$ erg $\rm cm^{-2}$ and $E_{p}=238$ keV;
GRB 160802A: 100--300 keV fluence $2.2\times10^{-5}$ erg $\rm cm^{-2}$ and $E_{p}=280$ keV;
GRB 160821A: 100--300 keV fluence $2.0\times10^{-4}$ erg $\rm cm^{-2}$ and $E_{p}=866$ keV;
GRB 160910A: 100--300 keV fluence $4.2\times10^{-6}$ erg $\rm cm^{-2}$ and $E_{p}=330$ keV \citep{2017arXiv170706595C})
to calculate the intrinsic peak energies and the isotropic gamma-ray energies for different
redshifts. By requiring that the bursts enter the $2\sigma$ region of the relation, we derive
$z\geq0.083$ for GRB 100826A, $z\geq0.082$ for GRB 110301A, $z\geq0.091$ for GRB 160106A,
$z\geq0.161$ for GRB 160325A, $z\geq0.105$ for GRB 160802A, $z\geq0.047$ for GRB 160821A,
and $z\geq0.272$ for GRB 160910A. Hereafter, the lower limits of redshifts are conservatively adopted.
We refer the reader to \cite{2014ApJ...783L..35D} for more details on the redshift estimation.

\begin{figure}
\begin{center}
\vskip-0.3in
\includegraphics[width=0.5\textwidth]{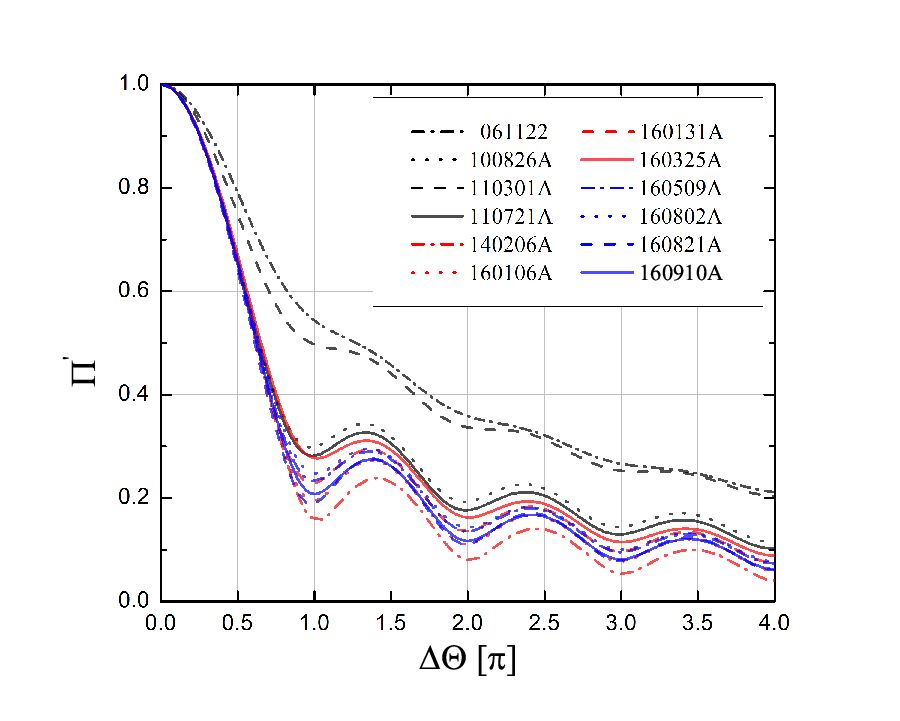}
\vskip-0.2in
\caption{Polarization degree $\Pi'$ as a function of the relative rotation angle $\Delta\Theta$
for different GRBs with known photon spectra and energy ranges.}
\label{fig:f1}
\vskip-0.3in
\end{center}
\end{figure}

\begin{figure*}
\begin{center}
\vskip-0.2in
\includegraphics[width=0.32\textwidth]{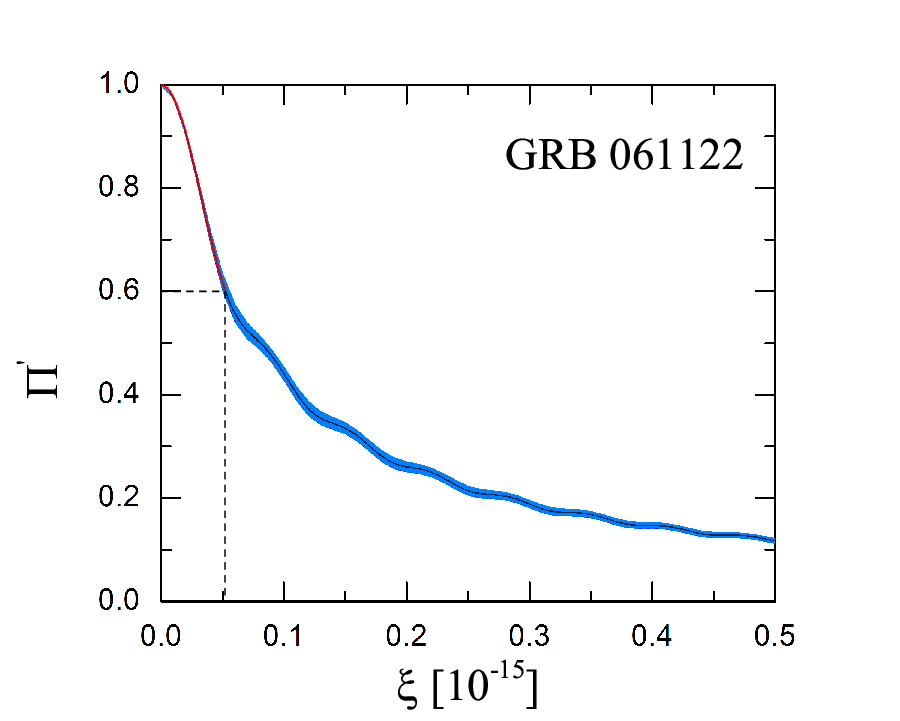}
\includegraphics[width=0.32\textwidth]{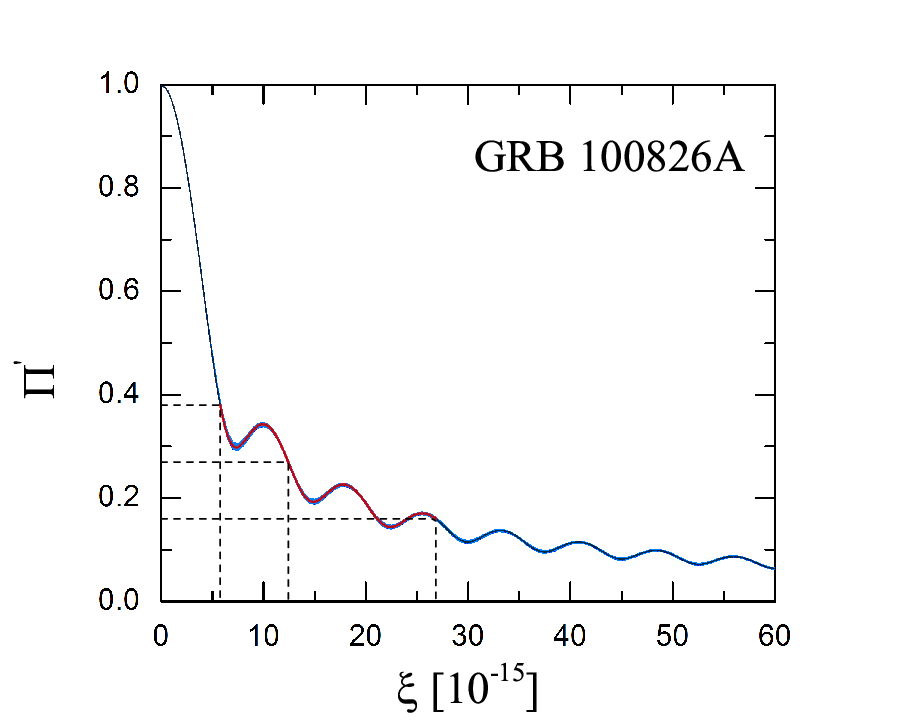}
\includegraphics[width=0.32\textwidth]{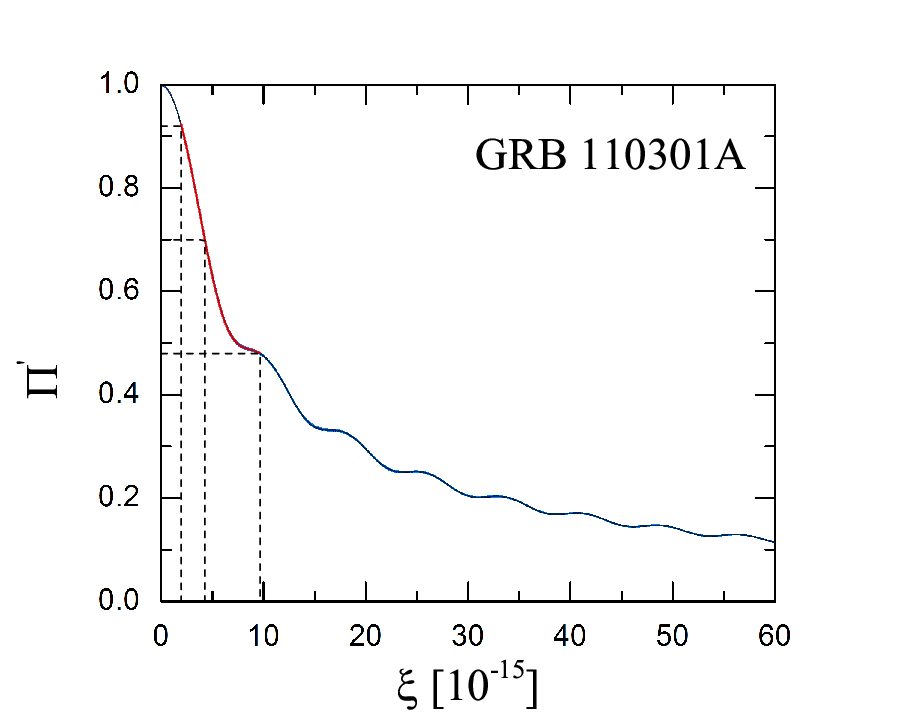}
\includegraphics[width=0.32\textwidth]{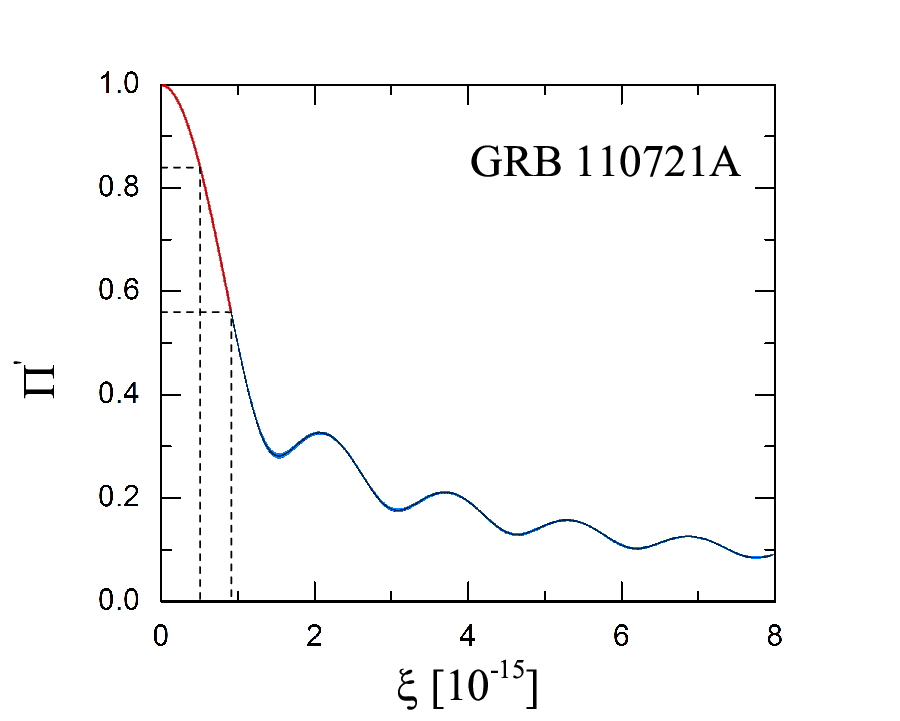}
\includegraphics[width=0.32\textwidth]{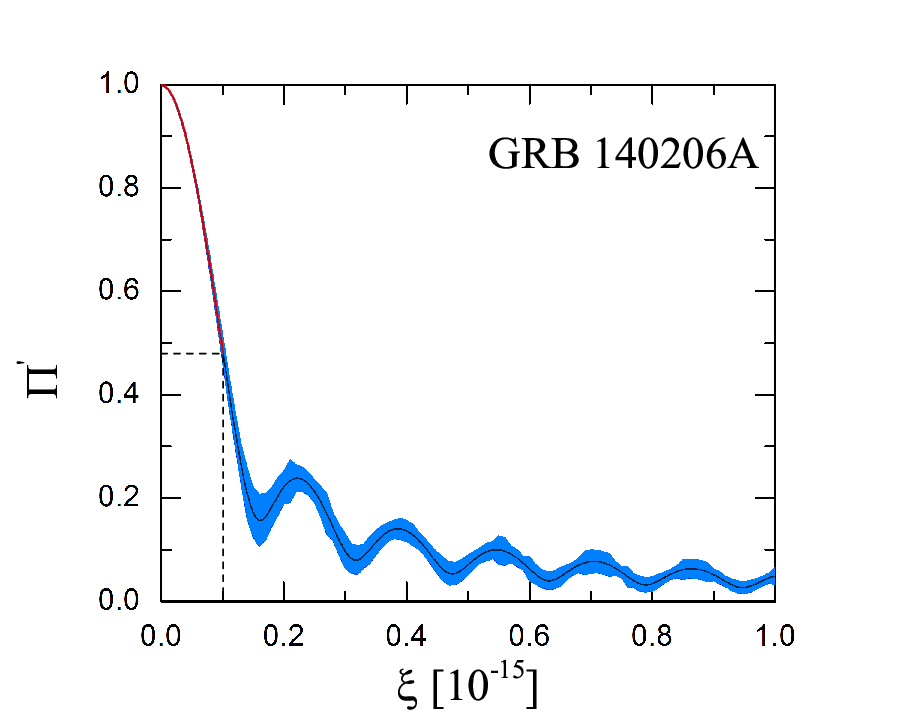}
\includegraphics[width=0.32\textwidth]{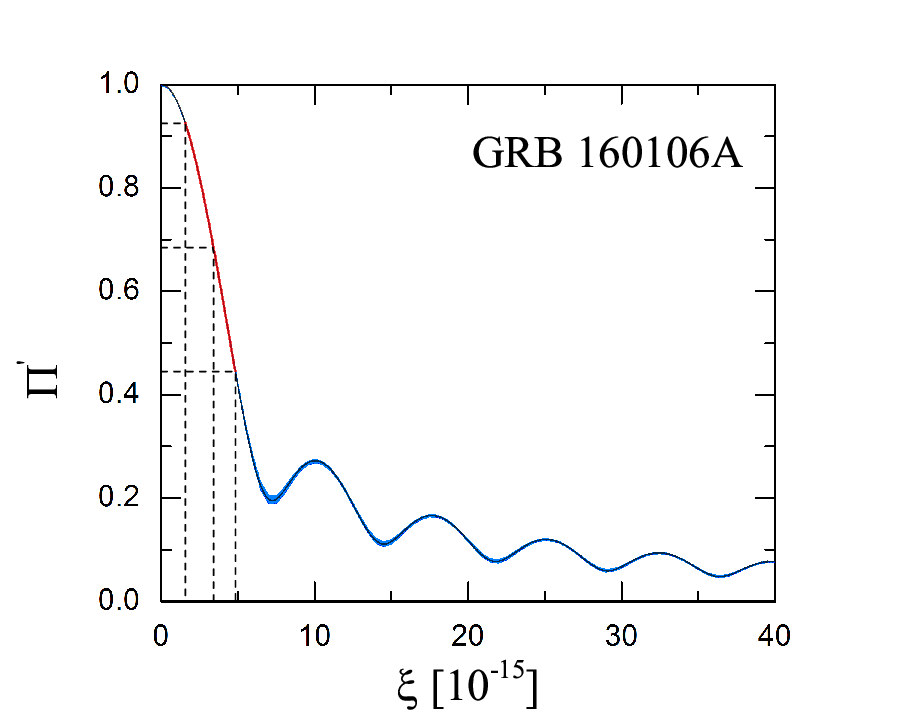}
\includegraphics[width=0.32\textwidth]{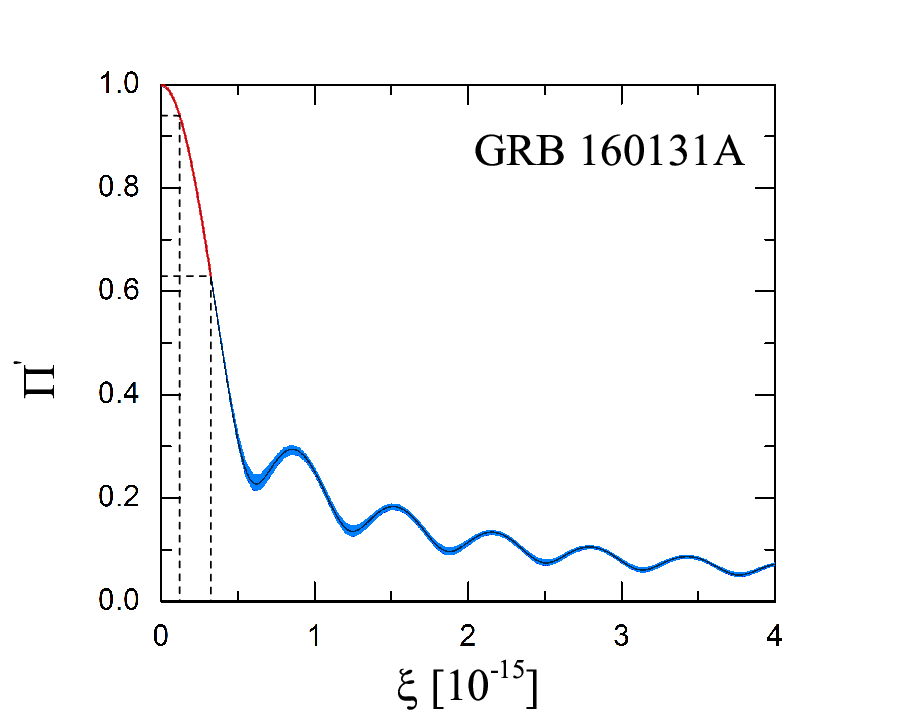}
\includegraphics[width=0.32\textwidth]{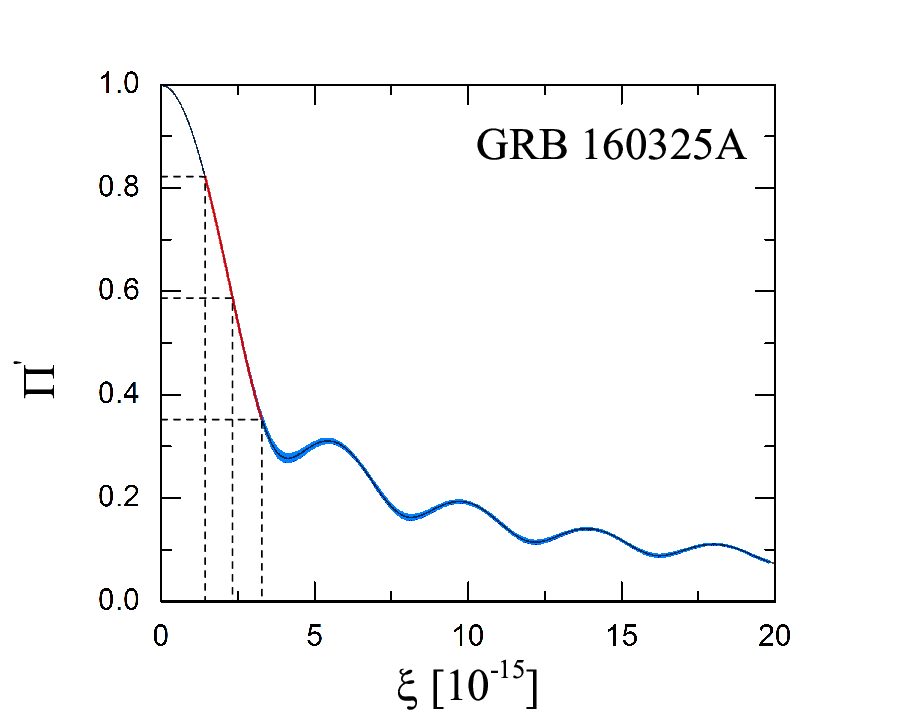}
\includegraphics[width=0.32\textwidth]{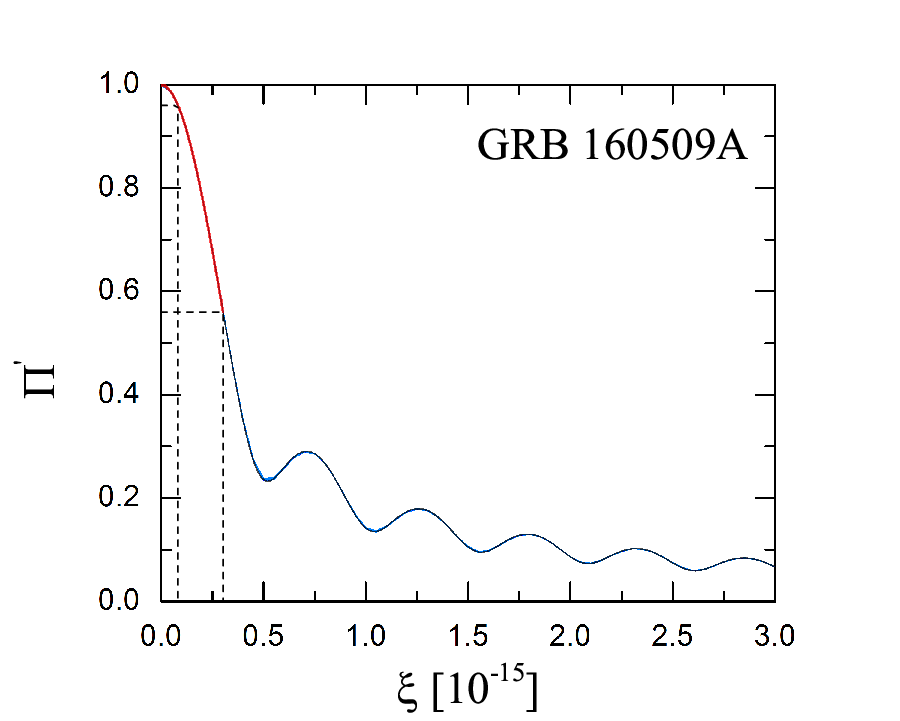}
\includegraphics[width=0.32\textwidth]{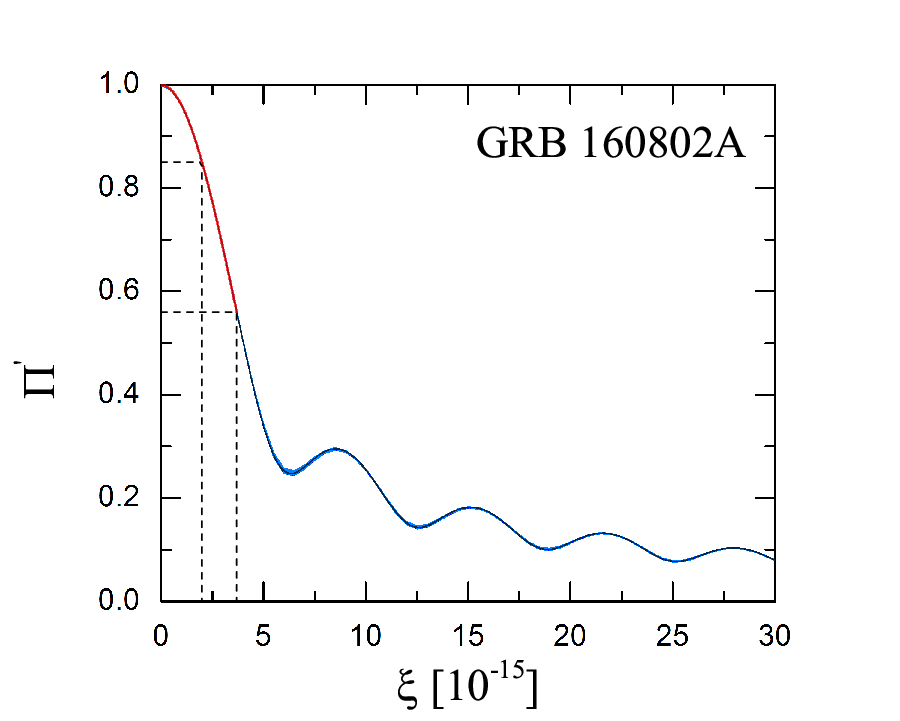}
\includegraphics[width=0.32\textwidth]{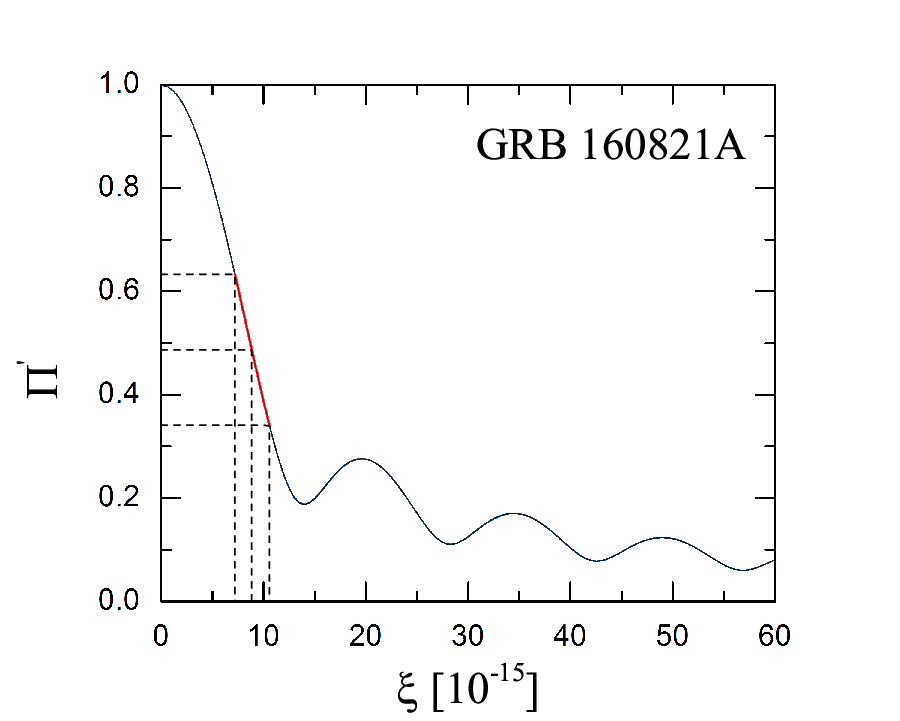}
\includegraphics[width=0.32\textwidth]{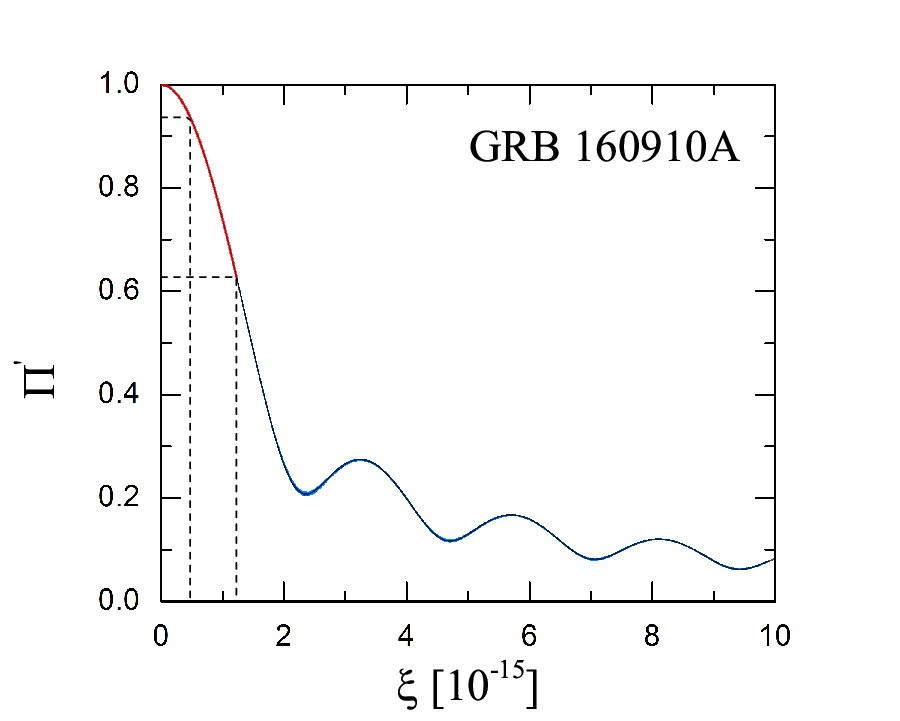}
\caption{Polarization degree $\Pi'$ as a function of the birefringence parameter $\xi$ for different GRBs.
The blue shaded areas, estimated by 1000 Monte Carlo simulations, represent the uncertainties due to
the observational uncertainties of spectral parameters.
The black solid curves represent the average of the Monte Carlo results.
The vertical and horizonal dashed lines mark the $\xi$ values that correspond to the observed polarization
degrees at $1\sigma$ C.L.. The red solid line regime corresponds to the $1\sigma$ constrained range of $\xi$.}
\label{fig:f2}
\vskip-0.2in
\end{center}
\end{figure*}

\section{Constraints on LIV}
\label{sec:results}
The photon spectrum of each GRB presented in Table~\ref{tab:1} can be well fitted by the Band function \citep{1993ApJ...413..281B}
or a power law with an exponential cutoff, i.e., $N(k)\propto k^{\alpha}\exp(-k/E_c)$. The third column of Table~\ref{tab:1}
is the energy range in which polarization is observed. With the known photon spectrum and energy range,
for any given relative rotation angle $\Delta\Theta$, we can numerically compute the maximum and minimum values of $I'(\varphi)$
according to Equation~(\ref{eq:observed-intensity}), and then calculate the polarization degree with Equation (\ref{eq:polarization2}).
The observed polarization degree $\Pi'$ as a function of $\Delta\Theta$ for different GRBs are shown in Figure~\ref{fig:f1}.
One can see from this plot that although the net polarization degree $\Pi'$ decreases rapidly with increasing $\Delta\Theta$
at $\Delta\Theta\leq\pi$, more than 60\% of the initial polarization degree can still be conserved at $\Delta\Theta\leq\pi/2$.
Note that with the chosen initial polarization pattern (the sum of $\delta$-functions)
and the typical spectral content of GRBs, the polarization measure is a multivalued function
for relative rotation angles above $\pi/2$. This is explicitly visualized in Figure~\ref{fig:f1}.
Thus, the polarization measure as a function of $\Delta\Theta$ can not be treated as an ``informative'' one
in the range $\Delta\Theta > \pi/2$ as being multivalued function for higher values of $\Delta\Theta$.
In other words, in practice, once the relative rotation angle exceeds $\sim\pi /2$ the polarization should be treated
as the lost one.
Additionally, due to the differences of spectral parameters and energy ranges, the polarization evolution
of GRBs are somewhat different (see Figure~\ref{fig:f1}). That is, the polarization evolution depends both on the photon spectrum
and the energy range (see \citealt{2016MNRAS.463..375L} for more explanations).

To show the resulting constraints on LIV more clearly, we also plot the polarization as a function of
the birefringence parameter $\xi$ for different GRBs in Figure~\ref{fig:f2}.
To explore the influence of the observational uncertainties of spectral parameters on the polarization evolution
and then on the test of LIV, we estimate uncertainties in the polarization evolution through 1000 Monte Carlo simulations
utilizing uncertainties in the spectral parameters.
The black solid curves show the polarization evolution from the average of the Monte Carlo simulations.
The blue shaded areas represent the uncertainties taking into account the spread of all the Monte Carlo simulation results,
which indicate that the uncertainties of spectral parameters do not have a significant impact on the polarization evolution.
With the observed polarization degrees, we can obtain strict limits on $\xi$ at 68\% C.L. for each burst,
which are displayed in column 10 of Table~\ref{tab:1}. These results represent sensitivities improved
by factors of 2 to 10-fold over existing bounds on a deviation from Lorentz invariance.

\section{Summary and discussion}
\label{sec:summary}
Violations of Lorentz invariance can produce vacuum birefringence, which leads to an energy-dependent rotation
of the polarization plane of linearly polarized photons. It has been confirmed that photons in the prompt emission
of some GRBs are highly polarized. The polarization measurements of GRBs have been widely used to constrain LIV.
Most of previous constraints were based on the assumption that the difference of rotation angles $\Delta\Theta$ of
polarization vectors between the highest and lowest energy photons is smaller than $\pi/2$, because it is believed
that $\Delta\Theta$ could not be too large when high polarization is observed, otherwise the net polarization
will be severely suppressed \citep{2012PhRvL.109x1104T,2013MNRAS.431.3550G,2014MNRAS.444.2776G}.
On the contrary, \cite{2016MNRAS.463..375L} found that more than 60\% of the initial polarization can still be conserved
at $\Delta\Theta\leq\pi/2$, by calculating the evolution of GRB polarization arising from the LIV effect.
But, because the polarization measure is a multivalued function for relative rotation angles exceeding $\pi/2$,
the polarization can not be treated as an ``informative'' one in the range $\Delta\Theta > \pi/2$.

In this work, we assemble recent measurements of linear polarization in the prompt gamma-ray emission of GRBs,
all of whom have high detection significance. For each GRB, the net polarization degree as a function of
$\Delta\Theta$ can be obtained by utilizing the method proposed in \cite{2016MNRAS.463..375L}.
We confirm that the net polarization is not significantly suppressed when $\Delta\Theta$ is approaching to $\pi/2$,
and that previous LIV tests by setting $\pi/2$ as the upper limit of $\Delta\Theta$ can be ameliorated.
With the observed polarization degrees, we place stringent limits on the birefringence parameter $\xi$, i.e., $10^{-14}-10^{-17}$,
improving previous limits on a deviation from Lorentz invariance by factors ranging from two to ten.
Additionally, we also demonstrate that our constraints are not greatly affected by uncertainties in the spectral parameters of GRBs.
At this point, it is interesting to make a comparison of recent achievements in sensitivity and
robustness of vacuum dispersion time-of-flight measurements versus polarization measurements.
Using the arrival-time difference of multi-GeV Fermi-LAT photons from GRB 090510A, \cite{2012PhRvL.108w1103N}
set the current best limit on the linear LIV energy scale, $M_{1}>525 M_{\rm Pl}$, which corresponds to $\xi<0.95\times10^{-3}$.
The latest result on time-of-flight analysis of multi-GeV signals from GRBs detected by Fermi-LAT is
$M_{1}>{\rm a\;few} \times10^{17}$ GeV \citep{2018arXiv180700189E}, which corresponds to $\xi<$ a few tens.
Compared to these time-of-flight constraints, polarization measurements obviously yield much more stringent limits.
As more and more GRB polarimeters (such as POLAR, TSUBAME, COSI, and GRAPE; \citealt{2017NewAR..76....1M}) enter service, it is reasonable to
expect that more GRBs with high polarization will be detected. More stringent limits on LIV can be expected
as this method discussed here is applied to larger number of GRBs with higher detection significance of polarization and higher redshifts.

\section*{Acknowledgements}
We thank the anonymous referee for constructive suggestions.
We also thank Prof. Xue-Feng Wu for helpful discussions.
This work is partially supported by the National Natural Science Foundation of China
(grant Nos. U1831122, 11603076, 11673068, and 11725314), the Youth Innovation Promotion
Association (2017366), the Key Research Program of Frontier Sciences (QYZDB-SSW-SYS005),
the Strategic Priority Research Program ``Multi-waveband gravitational wave Universe''
(grant No. XDB23000000) of the Chinese Academy of Sciences, and the ``333 Project''
and the Natural Science Foundation (grant No. BK20161096) of Jiangsu Province.






\bsp	
\label{lastpage}
\end{document}